# The Devil's Dung? Money as a mechanism of generalized reciprocity in human societies

Eduardo C. Ferraciolli[a], Francesco Renzini[b], Tanya V. Araújo[c], Flaminio Squazzoni[b]

[a] SOCIUS, Research Centre in Economic and Organisational Sociology, ISEG, Edifício Francesinhas 2, piso 2, Rua das Francesinhas, Lisbon, Portugal. Corresponding author: ec.ferraciolli@gmail.com

[b] BEHAVE, Department of Social and Political Sciences, University of Milan, Via Conservatorio 7, 20122 Milan, Italy.

[c] REM, Research in Economics and Mathematics, ISEG - Lisbon School of Economics and Management, University of Lisbon, Rua Miguel Lupi 20, 1249-078, Lisbon, Portugal

**Abstract:** St. Francis of Assisi (1181/82-1226) famously called money the devil's dung, and indeed money is often associated with greed, inequality, and corruption. Drawing on Nowak's five rules for the evolution of cooperation, we argue here that money promotes the formation of circuits of generalized reciprocity across human groups that are fundamental to social evolution. In an evolutionary tournament, we show that money exchange is an evolutionary stable strategy that promotes cooperation without relying on the cognitive demands of direct reciprocity or reputation mechanisms. However, we also find that excessive liquidity can be detrimental because it can distort the informational value of money as a signal of past cooperation, making defection more profitable. Our results suggest that, in addition to institutions that promoted trust and punishment, the emergence of institutions that regulated the money supply was key to maintaining generalized reciprocity within and across human groups.

# 1 Introduction

In an influential paper (Nowak, 2006), Nowak has proposed five mechanisms for the evolution of cooperation that are essential for the emergence of new levels of biological or cultural organization. The first, kin selection, applies to genetically related individuals: if a cooperator and the beneficiary of the cooperative action are close genetic relatives, altruistic acts can spread through kin relationships and outcompete defection (Axelrod & Hamilton, 1981; Dawkins, 1981). However, cooperation is also observed between genetically unrelated individuals through direct and indirect reciprocity (Okada, 2020; Redhead et al., 2024; Rossetti & Hilbe, 2024; Schmid et al., 2021). Direct reciprocity relies on personal memory: individuals cooperate only with those who cooperate with them. When repeated encounters between the same individuals are likely, it becomes easier to recall past actions of others, reducing defection and allowing mutually beneficial relationships to prevail. Indirect reciprocity, on the other hand, involves cooperating with others only if they have a positive reputation as previous cooperators. This expands the range of cooperative actions: agents $i$ can help other agents $j$ at time $t$ if they expect future benefits from agents $k$ at time $t + 1$, who have directly observed or indirectly knows about $i$'s cooperative efforts. This allows for the formation of broader cooperation circuits that do not depend on the individual memory of previous counterparts. Nowak's two additional mechanisms, network and group reciprocity, extend this analysis to relational structures and both inter- and intra-group selection.

However, human cooperation also occurs between individuals who are neither genetically related nor socially connected – that is, beyond the reach of personal memory, reputation, or shared membership in the same networks and groups (Robinson & Barker, 2017). Indeed, a remarkable feature of human societies is the extensive anonymous cooperation that takes place between complete strangers (Nowak, 2006; Bowles & Gintis, 2013; Simmel, 2011[1900]; Giddens, 1990; Powers et al., 2021). A prominent example is the economic division of labor (Giddens, 1990; Kocherlakota, 1998; Hayek & Bartley, 2000; Harwick, 2023; Nirjhor & Nakamaru, 2023). Individual $i$ would benefit from consuming good $x$ (e.g., rice), but lacks the resources or the ability to produce it independently. Instead, individual *i* relies on the costly efforts of a producer, individual *j*, with *i* 's benefits typically exceeding *j*'s costs. Individual *j*, in turn, depends on the costly efforts of other producers — often unrelated to individual *i* — to benefit from the use or consumption of good *y* (e.g., a mechanical tool). The refusal to bear the costs of production for the benefit of unrelated individuals, while at the same time reaping the rewards of the productive efforts of others, constitutes a defection from the expected cooperation, as it would jeopardize the division of labor.

Since the capacity to produce goods that satisfy increasingly sophisticated needs is unlikely to be found within one's own genetic pool or small village, cooperation among strangers becomes essential to gain access to goods and resources that would otherwise be inaccessible, leading to the emergence of more differentiated societies (Durkheim, 2023[1893]; Jaeggi et al., 2016; Robinson & Barker, 2017; Cooper & West, 2018; Shin et al., 2020; Powers et al., 2021; Fauvelle, 2025). However, this requires individuals to access information about unrelated others in order to assess their ability to engage in cooperative behavior (Pisor & Gurven, 2018; Shin et al., 2020).

In this paper, we propose that money can serve as a means of solving cooperation dilemmas, such as those observed in the context of the economic division of labor, by functioning as a "*token of delayed reciprocal altruism*" (Dawkins, 1981), thereby promoting the formation of circuits of generalized cooperation and reciprocity. Indeed, previous research has suggested that money functions as a public record-keeping device (Giddens, 1990; Ostroy, 1973; Ostroy, J. M. & Starr, R. M., 1990; Searle, 2005; Guala, 2020) or '*social memory*' mechanism (Kocherlakota, 1998; Hart, 2000) that facilitates the exchange of information between individuals who lack direct or indirect means of monitoring each other's past behavior (Kocherlakota, 1998; Camera et al., 2013; Fauvelle, 2025). Receiving monetary compensation for a cooperative action creates a tangible record of efforts expended to produce benefits for others. Possession of money thus signals an individual's history of cooperative behavior and the ability to reciprocate the costly efforts of others. This creates a dynamic circuit of cooperation in which individuals are incentivized to cooperate in exchange for money, which they can then use to incentivize further cooperation from others, such as the production of goods they need but cannot produce themselves.

Money can thus be understood as a key 'disembedding mechanism' (Giddens, 1990) that enables the expansion of coordinated social activities in time and space (Fauvelle, 2025). Indeed, the use of money is well documented across different historical contexts and societies, as shown by both archaeological artifacts and abstract systems of value accounting and debt (e.g., tally sticks, paper, gold, salt, cowrie shells) (Demps & Winterhalder, 2019). Despite great variation in other aspects of culture such as beliefs, social norms, or material practices, money-mediated exchange appears to be universal across many different social formations - especially as their size increases (Haour & Moffett, 2023; Fauvelle, 2025; Hudson, 2020; Kuijpers & Popa, 2021; Shin et al., 2020). For instance, recent archaeological studies report evidence of monetary-like exchange mechanisms taking place as early as the Late Pleistocene, with evidence from regions such as China and the Levant (Richter et al., 2011; Ridout-Sharpe, 2015; Singh & Glowacki, 2022; Xie et al., 2025).

The debate about the importance of generalized reciprocity for the evolution of human societies, and the evidence for the widespread, prehistoric diffusion of money, calls for a deeper analysis of the role that money may have played in shaping social evolution. To study this, we implemented a mechanism of money exchange for cooperative actions in an evolutionary game-theoretic simulation model where we assumed that each agent holds an initial number of tokens — goods or marks recorded in a reliable ledger — that have no intrinsic value and confer no evolutionary advantage by themselves. These collections constitute an agent's personal balance — countable, privately held, and nonperishable. We assumed that some agents would prefer to cooperate on the condition that they receive units of these balances in immediate exchange. These agents will expend the costly effort associated with cooperation only if their partner has a token and is willing to transfer that token in exchange for help. Specifically, they would be motivated to cooperate at a cost ($c$) at time $t$ in exchange for a token that they can later use to obtain cooperation from other agents — e.g., to secure their desired goods — at $t + 1$, thereby receiving a benefit $b > c > 0$. The ratio between benefits and costs of cooperation, $\frac{b}{c}$, is a key parameter of our model. Note that without receiving tokens, agents following this monetary exchange strategy would have no incentive to cooperate at all and would therefore defect.

We expect this strategy to maintain a population-level cooperation equilibrium if it is widely adopted (Camera et al., 2013; Bigoni et al., 2020). To test this hypothesis, we embedded agents following this monetary exchange strategy in a heterogeneous population under selective pressure similar to an evolutionary tournament (Axelrod, 2006; Nowak & Sigmund, 1992), which included four other types of competing agents in a indefinitely repeated helping game with random-matching (Axelrod & Hamilton, 1981; Bigoni et al., 2019; Camera et al., 2013; Nowak & Sigmund, 1998; Tkadlec et al., 2023). We initialized a balanced population of size $N = 500$ (Hauert & Doebeli, 2004; Nowak & Sigmund, 1992; Riolo et al., 2001), consisting of unconditional cooperators (always help, i.e., produce goods at cost $c$), defectors (never help, i.e., benefit only by receiving help from others), and both direct (help only agents who have not defected against me in previous interactions) and indirect reciprocators (help only agents with a positive image score), in addition to agents following the monetary exchange strategy. We implemented a relatively robust version of standard reciprocity strategies, in which direct reciprocators had no memory constraints, reputation scores were fully visible and free of noise, and agents could not make mistakes (Hilbe et al., 2017; Nowak & Sigmund, 1998; Panchanathan & Boyd, 2004). At each iteration of the tournament, agent strategies were probabilistically adjusted based on the in-round fitness of the surviving alternatives, following a roulette wheel selection process (Nowak & Sigmund, 1998; Tkadlec et al., 2023). As main results, we evaluated the evolutionary stability of each strategy and the overall cooperation rate at the population level.

This design served two purposes. First, it allowed us to test whether — and under what conditions — the monetary exchange strategy could be widely adopted even in a population where other conditional cooperation strategies were initially present. The evolutionary tournament in our model ensured that the eventual success of the monetary exchange strategy was not driven by a lack of plausible alternatives or by inefficiencies of the reciprocity strategies. Second, it allowed us to identify the specific aspects of the monetary exchange strategy that distinguish it from reciprocity-based strategies and directly influence the evolutionary trajectories of the population. For instance, the so-called *second-order defection problem* (Nowak & Sigmund, 1998; Okada, 2020) poses significant challenges to the ability of reputation-based mechanisms to sustain cooperation in the long run. In particular, even relatively complex indirect reciprocity strategies that successfully eliminate defectors can become vulnerable to invasion by unconditional cooperators, which ultimately reintroduces an evolutionary advantage for mutations that promote defection. This determines recurrent cycles of defection and cooperation. For cooperation to be evolutionarily stable, strategies are needed that — unlike reciprocity strategies — eliminate both unconditional defectors *and* cooperators from mixed populations. Money fulfills exactly this function.

## 2  The Model

We consider a population of $N$ agents ($N = 500$) playing an indefinitely repeated helping game with random matching. At each time step, each agent is paired with a randomly selected partner and must decide whether to cooperate or defect. Cooperation involves providing a benefit $b$ to the partner (referred to as 'help' in the game) while incurring a cost $c$ (with $b > c > 0$), while defection involves withholding help, providing no benefit and incurring no cost (Nowak, 2006). We define an agent's fitness as the difference between benefits received and costs paid in a given round. For most agents, in-round fitness will simply be either $b - c$ (if they provide and receive help), $b$ (if they only receive help), $-c$ (if they only provide help), or zero (if no help is provided or received). However, due to stochasticity, multiple agents may occasionally decide whether to help the same counterpart. In such cases, if these multiple agents provide help, the recipient's fitness may exceed $b$ — in particular, it can reach $hb$, if all $h > 1$ helping agents cooperate and the recipient does not provide help to its own counterpart. For simplicity, we fixed $c = 1$, and we determined $b$ by exploring a wide range of benefit-cost ratios, going from exceptionally low ($\frac{b}{c} = 1.25$) to extremely high ($\frac{b}{c} = 100$).

In the evolutionary tournament, each agent follows one of five different strategies, initially equally distributed in the population (48): *(i) unconditional cooperation*, i.e., they always help their randomly matched partners; *(ii) defection*, i.e., they never help their current partners; *(iii) direct reciprocity*, i.e., they help their current partners as long as the partner has not previously defected against them; *(iv) indirect reciprocity*, i.e., they help their current partners as long as their reputation score as cooperators is positive; and *(v) monetary-exchange*, i.e., they help their partners as long as the partners can transfer one unit of their token balances in exchange for help.

We implemented a relatively robust version of each reciprocity strategy. All agents have a memory "blacklist" that tracks previous past defectors. Direct reciprocators only cooperate with partners who are not in the list; otherwise, they punish their partners by defecting, removing them from their memory, and becoming willing to cooperate again (Hilbe et al., 2017). All agents also have a personal score that tracks their reputation. Indirect reciprocators cooperate only if their partner's score is positive, otherwise they defect (Nowak & Sigmund, 1998; Schmid et al., 2021). If an agent cooperates, the reputation score is increased by one unit; if the agent defects, the score is decreased by one unit. Agents are initialized with moderately positive reputation scores. We do not implement any memory constraints, and all reputation scores are perfectly visible and noise-free (Panchanathan & Boyd, 2004).

The initial amount of money held by each agent is derived from a *liquidity* parameter as follows. For a liquidity value of 1, each agent is initialized with one unit of money; for values greater than 1, each agent receives the corresponding units of money; for values lower than 1 (e.g., 0.5), only the corresponding percentage of agents (e.g., 50%) are randomly initialized with one unit of money.

The importance of memory, reputation scores, and money balances varies depending on the strategy. For instance, agents following the monetary exchange strategy have a memory list, but they do not use it to decide whether to provide help. Similarly, direct reciprocators may have money balances, but their decisions are independent of the monetary balances of others.

Figure 1 visually summarizes the main endogenous effects of cooperation and defection on fitness and such important agent-level variables. Regardless of the strategy an agent follows, cooperative actions always reduce the fitness of the helping agent by $c = 1$. At the same time, cooperation increases the agent's reputation score by one unit. At the same time, the recipient receives a fitness increase of $+b$, where $b$ is determined from the benefit-to-cost ratio as follows: $\frac{b}{c} \times c = b$. If the helping agent follows the monetary exchange strategy, the helping agent's token balance increases by one unit and the recipient's token balance decreases by one unit. As previously discussed and summarized in the table in the middle of Figure 1, this monetary exchange can only occur if recipients have at least one token in their balance at the time of the interaction.

Defection does not affect the fitness of either the helping agent or the recipient. It does, however, affect the agents' memory. When an agent defects, the partner adds the defector to their memory list (if they were not already present). Conversely, if the defecting agent follows a direct reciprocity strategy, this action may 'free' a previously stored agent from the list, effectively removing them from memory. This means that direct reciprocators do not follow a grim trigger strategy, such as 'if you defect once, I will defect forever' (J. W. Friedman, 1971).

Defection also has no effect on token balances. If an agent following the monetary exchange strategy refuses to help a recipient because the recipient lacks tokens to exchange, the agent's token holdings remain unchanged. Here, we assume that there is no retaliation — recipients who do not receive help do not resort to stealing or damaging the wealth of the refusing helping agent. Finally, defection always implies a reputation loss by decreasing the reputation scores by one unit.

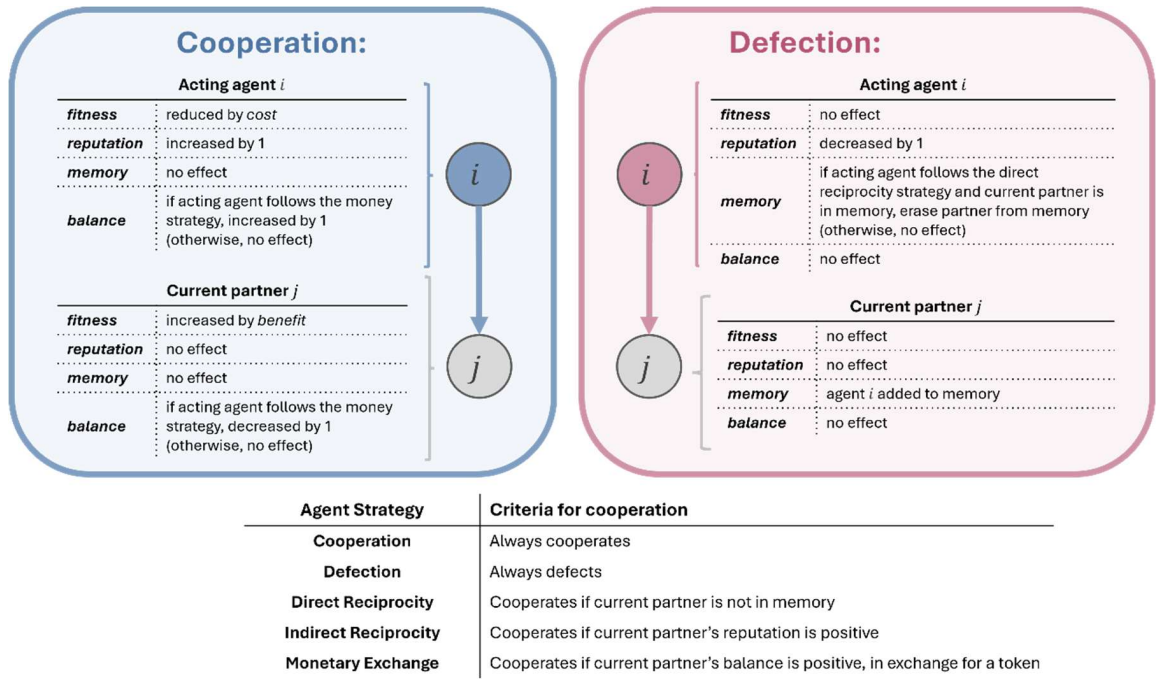

Figure 1. Summary of the endogenous effects of cooperation and defection on key agent-level variables, along with a brief description of the conditions under which each strategy chooses to cooperate.

At each step, each agent is randomly paired with a partner and decides whether to provide help based on their current strategy and either theirs or their partner's relevant variables, such as personal memory, token balance, or reputation score. The fitness of the agents is evaluated after everyone has acted. Subsequently, a randomly selected agent may adopt a new strategy with a probability proportional to the relative fitness of the surviving alternatives in the current round, as in a roulette wheel selection process (Nowak & Sigmund, 1998; Tkadlec et al., 2023). Since $c = 1$, we increase the fitness of all agents by +1 to ensure non-negative values for probability calculations (Nowak & Sigmund, 1998). The roulette wheel selection process allows us to keep the population size constant while introducing random noise into the selection mechanism. For simplicity, the fitness values of the agents are reset at the end of each step.

We rely on two output metrics to evaluate the success of each strategy and its consequences. First, we track the number of agents adopting each of the five strategies over time. Second, we evaluate the overall cooperation rate, defined as the number of helping actions (i.e., a cooperation counter) divided by the total number of interactions taking place during each time step. The cooperation counter is reset at the end of each time step, after saving the previous value. To optimize memory usage without losing essential information, simulation data were stored locally at intervals of 250 time steps.

We examined the following 9 parameter values for $\frac{b}{c}$ ratios: $\{1.25, 1.5, 2, 3, 5, 10, 20, 50, 100\}$. We also varied the liquidity parameter across 13 possible values: $\{0.25, 0.5, 1, 2, 5, 10, 15, 20, 35, 50, 100, 250, 500\}$, generating 117 possible parameter combinations that could influence evolutionary selection and overall cooperation rates. We set the initial reputation score to 1 for all agents in each scenario and allowed it to evolve endogenously. For each parameter combination in each scenario, we ran 100 parallel repetitions. All simulations were run for 30000 time steps, ensuring sufficient duration to capture the long-term equilibrium effects of agent interactions. For instance, this means that each agent is selected on average 60 times to potentially update its strategy. The Supplementary Information (SI) file presents a benchmark scenario without the monetary strategy (evaluating the evolutionary performance of reciprocity-based strategies when competing exclusively with unconditional cooperators and defectors), as well as further robustness checks for selected parameter combinations.

**Model Documentation, Data Analysis and Supplementary Materials**

The model was implemented in both NetLogo (version 6.4.0) (Wilensky, 1999) and C++ (C++17 standard). To ensure full reproducibility, the SI file also includes the model pseudocode and a detailed execution flowchart. The simulated data were analyzed using R (version 4.3.3). The Supplementary Information file, the open source code for both implementations, the data used to generate the figures, and the data analysis scripts are available at:

https://gitfront.io/r/anonymized-submission/J6gSBcvfgcsm/money-generalized-cooperation-reciprocity/

## 3 Results

Figure 2 shows that the monetary exchange strategy (in green) thrives in all environments, regardless of whether the returns to cooperation are high or low. This pattern holds over a wide range of liquidity values. Even modest initial liquidity (e.g., 0.25, where only 25% of agents randomly receive 1 token during initialization) allows the monetary exchange strategy to outperform all others, eventually becoming dominant across a wide range of benefit-cost ratios. This is a unique property of the monetary exchange strategy. Indeed, our benchmark in the SI file shows that under these tournament settings, a mix of reciprocity strategies can become dominant only when the benefit-to-cost ratio is sufficiently high (Figure S2 in the SI file).

When the monetary exchange strategy is widely adopted, population-level cooperation rates tend to show low variability, as indicated by the increasingly narrow interquartile ranges, thereby stabilizing cooperation around a particular equilibrium cooperation level. However, an equilibrium of full cooperation does not always emerge when the monetary exchange strategy proliferates because, depending on the initial liquidity levels, agents may lack sufficient balances to compensate their randomly matched partners for cooperation. Consequently, the stabilization of cooperation by the money mechanism only generates higher payoffs for the entire population when liquidity levels are sufficiently high (Figure 2).

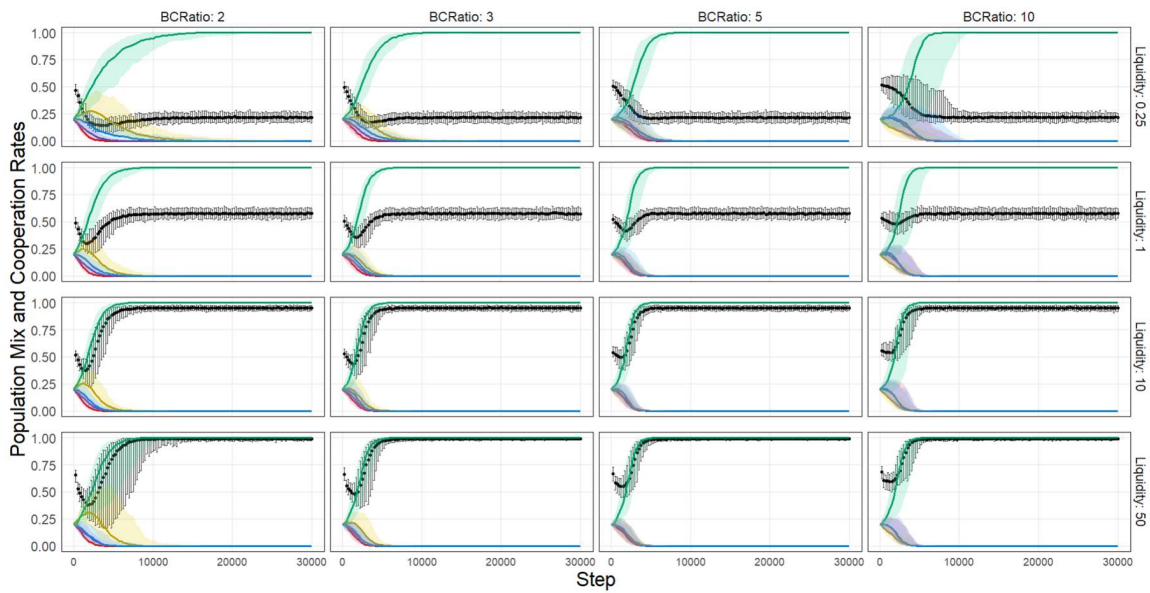

**Figure 2. Evolutionary dynamics of population composition and cooperation in the tournament model under selected benefit-to-cost ratios and liquidity levels.** Colored lines represent the median population share of each surviving strategy over time, with shaded ribbon bands indicating interquartile ranges (IQR). Specifically, green denotes agents following the monetary exchange strategy, yellow defectors, red unconditional cooperators, light

blue indirect reciprocators, and purple direct reciprocators. The black line and associated error bars show the median and IQR of the cooperation rate in the population.

Figure 3 better illustrates how initial liquidity levels affect overall cooperation rates when returns to cooperation are low, as a selected case. Under these conditions, the monetary exchange strategy tends to dominate all other strategies, even when initial liquidity is scarce (Figure 2). In such cases, final cooperation rates are low because there are few money units in circulation, and cooperative actions occur only when a money user meets a counterpart with positive balances. As liquidity increases, cooperation rates rise according to a non-linear relationship, reaching a saturation point at 100%. At moderately high liquidity levels, the monetary exchange strategy outperforms other strategies while maintaining very high cooperation rates, as most agents have sufficient money balances to exchange in most encounters. However, at extremely high initial liquidity levels, cooperation is fragile: the stability of the monetary exchange strategy may be compromised, and defectors may proliferate.

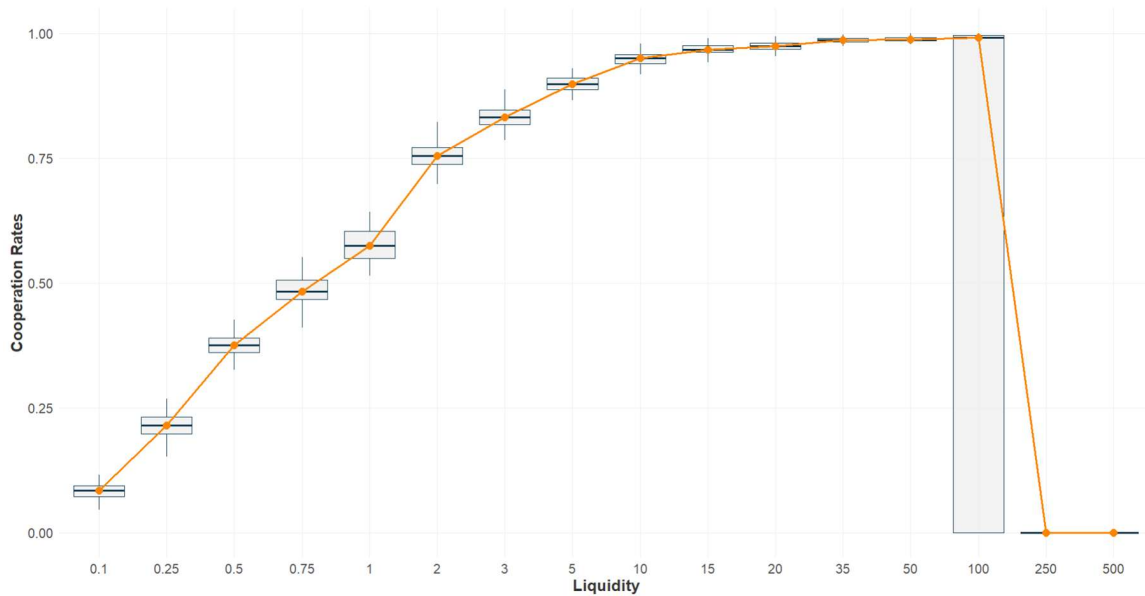

**Figure 3. Cooperation rates across different levels of initial liquidity at the 30000th simulation step. The orange line connects the medians of the distributions of cooperation rates across liquidity levels, highlighting the overall trend and the excess liquidity effect. Here, the benefit-cost ratio was held constant at 2.**

To better understand these dynamics, consider two extremes. With zero initial liquidity, agents following the monetary exchange strategy always defect: there is no incentive for them to cooperate since no agent in the population holds money. On the other hand, with infinite liquidity, these agents behave as unconditional cooperators: they cooperate in every interaction, since each partner always has sufficient balances to exchange for help. In this case, even defectors continuously benefit from the cooperation of these agents at no cost.

At intermediate levels of liquidity, the use of money becomes a discriminator strategy, similar to the famous 'green beard' effect (Dawkins, 1981; Hamilton, 1964). Through multiple transactions, tokens eventually become concentrated among agents who follow the monetary exchange strategy because they are the only ones who demand them. As a result, these agents end up cooperating exclusively with each other, creating a money-induced cooperative circuit that expands as the monetary exchange strategy spreads. At these intermediary liquidity levels, cooperation actions within the circuit increase nonlinearly with liquidity: as tokens become more widely available, agents are more likely to have sufficient balances to exchange for help, leading to higher cooperation rates. However, when liquidity is extremely high, a tipping point is reached where the discriminatory effect ceases to work, and defectors may manage to stay in the money-induced cooperation circuit. With excessive liquidity, token ownership no longer reflects previous cooperative actions, allowing defectors to continuously obtain cooperation without reciprocating. Consequently, the monetary exchange strategy loses its evolutionary advantage in favor of costless defection.

Figure 4 provides a comprehensive overview of how different combinations of benefit-cost ratios and liquidity levels affect equilibrium cooperation rates and population composition. Figure 4 shows that this 'excess liquidity effect' depends on the returns to cooperation. At high benefit-cost ratios, even substantial levels of liquidity do not undermine cooperation. Conversely, at low benefit-cost ratios, high levels of liquidity are detrimental to cooperation (Figure 3), as defectors proliferate. This highlights the need to ensure an adequate level of money balances in the system to sustain cooperation when its benefits are minimal.

| Liquidity \ Benefit-Cost Ratio | 1.25 | 1.5 | 2 | 3 | 5 | 10 | 20 | 50 | 100 |
|---|---|---|---|---|---|---|---|---|---|
| 0.25 | 0.18 (0.05) | 0.21 (0.03) | 0.21 (0.02) | 0.22 (0.02) | 0.22 (0.03) | 0.22 (0.02) | 0.22 (0.02) | 0.28 (0.22) | 0.27 (0.18) |
| 0.5 | 0.36 (0.04) | 0.38 (0.03) | 0.38 (0.02) | 0.38 (0.03) | 0.37 (0.03) | 0.38 (0.03) | 0.38 (0.05) | 0.38 (0.07) | 0.40 (0.12) |
| 1 | 0.57 (0.03) | 0.58 (0.02) | 0.58 (0.03) | 0.58 (0.03) | 0.58 (0.03) | 0.58 (0.03) | 0.58 (0.03) | 0.59 (0.05) | 0.58 (0.04) |
| 2 | 0.76 (0.03) | 0.76 (0.02) | 0.76 (0.02) | 0.75 (0.02) | 0.76 (0.02) | 0.76 (0.02) | 0.76 (0.03) | 0.76 (0.04) | 0.76 (0.02) |
| 5 | 0.89 (0.08) | 0.90 (0.02) | 0.90 (0.02) | 0.90 (0.02) | 0.90 (0.02) | 0.90 (0.02) | 0.90 (0.02) | 0.90 (0.02) | 0.90 (0.02) |
| 10 | 0.92 (0.16) | 0.95 (0.01) | 0.95 (0.01) | 0.95 (0.01) | 0.95 (0.01) | 0.95 (0.01) | 0.95 (0.01) | 0.95 (0.01) | 0.95 (0.01) |
| 15 | 0.79 (0.36) | 0.97 (0.01) | 0.97 (0.01) | 0.97 (0.01) | 0.97 (0.01) | 0.97 (0.01) | 0.97 (0.01) | 0.97 (0.01) | 0.97 (0.01) |
| 20 | 0.69 (0.44) | 0.96 (0.14) | 0.98 (0.01) | 0.98 (0.01) | 0.98 (0.01) | 0.97 (0.01) | 0.98 (0.01) | 0.97 (0.01) | 0.97 (0.01) |
| 35 | 0.22 (0.40) | 0.88 (0.31) | 0.99 (0.01) | 0.99 (0.01) | 0.99 (0.01) | 0.99 (0.01) | 0.99 (0.01) | 0.99 (0.01) | 0.99 (0.01) |
| 50 | 0.00 (0.00) | 0.59 (0.49) | 0.96 (0.17) | 0.99 (0.01) | 0.99 (0.01) | 0.99 (0.01) | 0.99 (0.01) | 0.99 (0.01) | 0.99 (0.01) |
| 100 | 0.00 (0.00) | 0.01 (0.10) | 0.57 (0.49) | 0.98 (0.10) | 0.99 (0.01) | 0.99 (0.00) | 0.99 (0.00) | 0.99 (0.00) | 0.99 (0.00) |
| 250 | 0.00 (0.00) | 0.00 (0.00) | 0.01 (0.10) | 0.32 (0.47) | 0.98 (0.14) | 0.98 (0.14) | 1.00 (0.00) | 1.00 (0.00) | 1.00 (0.00) |
| 500 | 0.00 (0.00) | 0.00 (0.00) | 0.00 (0.00) | 0.00 (0.00) | 0.46 (0.50) | 0.99 (0.10) | 0.99 (0.04) | 1.00 (0.01) | 1.00 (0.01) |

**Figure 4.** Cooperation rates and dominant strategy in the population at the end of the simulation (30000th step) for broader combinations of benefit-to-cost ratios and liquidity values. Larger numbers indicate the mean cooperation rate for each combination of benefit-to-cost ratio and liquidity level, with smaller numbers in parentheses indicating the standard deviations over 100 independent runs. Cell colors represent the dominant strategy: green for money and yellow for defection. A transparency gradient visually illustrates mean cooperation rates across the parameter space, with darker shades indicating higher cooperation.

The success of the monetary exchange strategy depends on the emergence of the money-induced exchange circuit, which expands over time as tokens become increasingly concentrated within and circulate exclusively among agents who condition their cooperation on receiving a token in return. Other agents can only gain access to this circuit by adopting the monetary strategy themselves — fueling its expansion by reinforcing a self-reinforcing lock-in effect. By adhering to this simple conditional cooperation rule, agents avoid the costs of cooperating with those outside the circuit and pay a fitness cost only to help and support other circuit members.

This results in strong *positive assortativity* (Nax & Rigos, 2016; Iyer & Killingback, 2020), creating an impermeable boundary between an 'out-group' of non-money users and an 'in-group' of money users, where cooperative interactions occur. Interestingly, this separation emerges even without requiring agents to actively recognize their fellow members or signal their affiliation through costly or covert actions (e.g. Gambetta, 2011; Smaldino et al., 2018). Instead, a simple litmus test suffices: whether the recipient of one's costly cooperative action possesses a token that can later be used to incentivize others to cooperate. This litmus test excludes not only defectors, but also unconditional cooperators and other conditional cooperation strategies from the circuit (Figures 2 and 4).

This means that unconditional cooperators are not 'neutral mutants' of money users (Okada, 2020). In a population consisting only of unconditional cooperators and money users, the latter will eventually stop helping the former while continuing to receive their help, leading to the negative selection of unconditional cooperators (Figure 5). This prevents 'defection cascades' (Nowak & Sigmund, 1998; Okada, 2020) and ensures that a population composed entirely of agents following the monetary exchange strategy remains resistant to late-stage invasions by defectors or defection-promoting mutations. In the SI, we show that forcibly converting half of the population of agents following the monetary exchange strategy into defectors, once it has become dominant, has no long-term consequences for the system (see Figure S3 in the SI). The cooperation rate quickly returns to its pre-invasion level, demonstrating the resilience of the monetary exchange circuit in sustaining cooperation. Moreover, the SI shows that the money-induced circuit remains robust even when invaded by unconditional cooperators or a mix of unconditional cooperators and defectors (Figures S4 and S5 of the SI file).

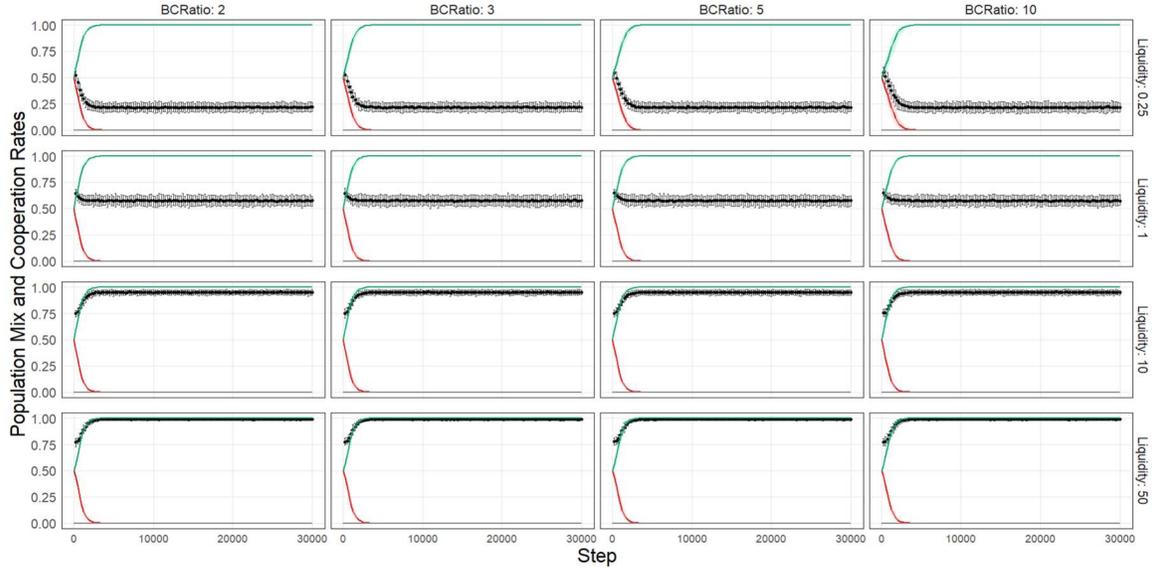

**Figure 5.** Evolutionary dynamics of population composition and cooperation in the tournament model under selected benefit-to-cost ratios and liquidity levels, starting from a population composed exclusively of unconditional cooperators (red) and agents following the monetary exchange strategy (green). The lines represent the median population share of each surviving strategy over time, with shaded ribbon bands indicating interquartile ranges (IQRs). Note that the ribbons are barely visible due to minimal variability across simulation runs, as the monetary exchange strategy quickly outcompetes unconditional cooperators and eliminates them from the population. The black line and associated error bands show the median and IQR of the overall cooperation rate.

More nuanced interaction effects reveal how the monetary circuit emerges and thrives even in the presence of competing conditional cooperation strategies, such as the reciprocity-based ones examined in our evolutionary tournament. Even at very low levels of liquidity (Figures 2 and 4), agents following the monetary exchange strategy effectively act as unconditional cooperators in the early stages of the simulation. This allows them to quickly build a strong positive reputation among indirect reciprocators and avoid being 'blacklisted' by direct reciprocators. As a result, they consistently receive help from both direct and indirect reciprocators, in addition to unconditional cooperators.

However, this equilibrium is only temporary and collapses once the token balances of competing strategies are completely exhausted. The rate at which this happens is inversely proportional to the total liquidity. When depletion occurs, agents following the monetary exchange strategy stop cooperating with reciprocity-based strategies while continuing to receive help from both direct and indirect reciprocators. This is because non-money users deplete their token balances before the reputation of money users deteriorates due to defection. As a result, money users temporarily exploit reciprocity-based strategies in a manner similar to defectors. However, unlike defectors, they continue to cooperate with each other within the monetary circuit. This gives the monetary exchange strategy a strong evolutionary advantage, allowing it to 'piggyback' on other conditional cooperation strategies due to a temporal mismatch in how these strategies respond to changes in the competitive context. While this effect is observed for our specific operationalizations of direct and indirect reciprocity, it suggests a broader principle: as long as any reciprocity-based strategy takes sufficient time to recognize money users as defectors, the monetary exchange strategy can proliferate by exploiting this delay.

As a robustness check, Figures S6 – S9 in the SI include alternative versions of Figures 2 and 4, where the initial population exclusively consists of unconditional cooperators, defectors, and agents following the monetary exchange strategy. These tests confirm that the mechanisms and results described here remain consistent, even when the monetary exchange strategy does not piggyback on reciprocity strategies during the early stages of the formation of the money-induced cooperation circuit. In short, the monetary exchange strategy has a strong evolutionary advantage due to its ability to establish and maintain an expanding circuit in which cooperation occurs exclusively within it.

Figure 6 provides a distilled visualization of this mechanism in a simplified setting. In panel A, we observe the early stages of the tournament. Cooperation between agents is represented by directed links. In green we see agents following the monetary exchange strategy, which initially cooperate (green arrows) with other strategies (in gray) because they have a positive money balance $B$. The other strategies — direct and indirect reciprocators, unconditional cooperators, and defectors — may or may not cooperate with each other or with the monetary exchange strategies, as indicated by dashed gray arrows. Over time, money balances become increasingly concentrated among money users (Panel B), who start forming a cooperative circuit that includes a growing number of agents who adopt this strategy. These agents avoid cooperating with others who have exhausted their balances (marked in orange). While the remaining strategies can still cooperate with each other and with monetary exchange users, cooperation becomes relatively more common within the circuit. Outside the circuit, defection increases as strategies retaliate against each other.

This dynamic continues in panel C, where more agents switch to the monetary exchange strategy, and tokens become completely concentrated within the circuit. At this stage, there is no cooperation between circuit members and agents from other strategies who have exhausted their tokens, although the latter may still cooperate with each other. Finally, at equilibrium, all agents adopt the monetary exchange strategy. As discussed above, the only threat to the stability of this system is an excessive initial supply of tokens, which allows defectors to maintain access to the money-induced circuit and eventually undermine it from within.

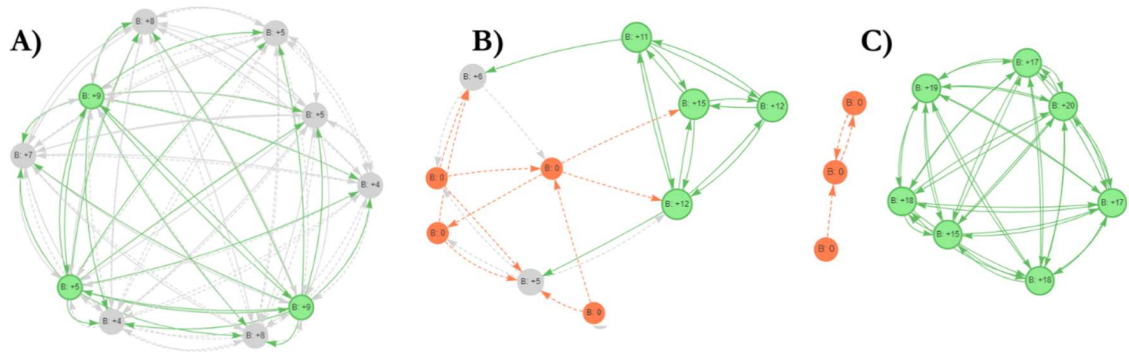

**Figure 6. A simplified representation of the evolution of the money-induced cooperation circuit.** Agents following the monetary exchange strategy are shown in green, and their number increases over time due to their evolutionary success. Other strategies are depicted in gray but turn orange when their balance ($B$) reaches zero. Cooperation is represented by directed links. Green arrows indicate cooperation patterns of agents following the monetary exchange strategy, while dashed arrows indicate that other strategies may or may not cooperate with each other or with money users.

## 4 Discussion

Our results show that the monetary exchange strategy can consistently proliferate in a mixed population under evolutionary pressure by establishing a self-sustaining circuit of money-contingent cooperation. This occurs even in the presence of other conditional cooperation strategies, while also enhancing and stabilizing population-level cooperation across a wide range of benefit-cost ratios and liquidity levels. At the same time, we also found that an excessive liquidity effect can undermine the evolutionary stability of the money mechanism when the benefits of cooperation are low. The role of liquidity in the success of the monetary exchange strategy is analogous to key parameters in other evolutionary models of cooperation, such as genetic relatedness, the probability of repeated interactions, or the visibility of reputation (Nowak, 2006).

In "Five Rules for the Evolution of Cooperation", Nowak writes: "[…] Direct reciprocity is like a barter economy based on the immediate exchange of goods, whereas indirect reciprocity resembles the invention of money. The money that fuels the engines of indirect reciprocity is reputation" (Nowak, 2006, p. 3). Our model suggests that money *itself* should be seen as a generalized reciprocity mechanism: a special form of token-mediated reciprocity providing a kind of *sixth rule* for the evolution of cooperation. This simple rule consists of *cooperating with agents who have positive money balances, not cooperating otherwise, and cooperating only in exchange for a transfer of money*. This guarantees the formation of the self-policing circuits necessary to promote generalized reciprocity. Provided that liquidity remains within reasonable limits, agents will only cooperate with those who have previously cooperated with others and can prove it by exchanging a money token.

Unlike the individual links in direct reciprocity, or the within-group cycles of cooperation sustained by reputation in indirect reciprocity, money creates a stable, self-sustaining circuit of cooperation without requiring individuals to keep track of personal interactions or rely on any form of group-level communication or gossip (Nowak, 2006). Each transfer of money between pairs of agents corresponds to a cooperative action in the opposite direction, linking individuals in a long chain of decentralized, anonymous cooperation. In this sense, money functions as a form of 'social memory' (Hart, 2000; Kocherlakota, 1998), providing information about past cooperative actions within a given population.

This seems to reflect the way reputation facilitates the generalization of cooperation in indirect reciprocity. However, unlike in reputation-based systems, the monetary circuit can police itself. As tokens held by other agents migrate to agents following the monetary exchange strategy through multiple transactions, both unconditional cooperators and defectors are gradually excluded from the circuit of money-induced cooperation as their token supply is depleted. This dynamic provides a novel solution to the first and second-order free-rider problems (Okada, 2020), making the monetary exchange strategy resilient to defection cascades that typically weaken reciprocity-based mechanisms.

The success of the monetary mechanism is also driven by two key differences between money and indirect reciprocity. The first is *interdependence*. In reputation-based systems, only the reputation score of the agent in the helping role is updated after cooperation or defection, while the recipient's score remains unaffected (Nowak & Sigmund, 1998). This is because reputation ultimately exists in the 'eyes of the beholder' (Nowak & Sigmund, 2005) and subsequently reaches other agents through gossip or other forms of nonverbal communication. In contrast, in a monetary exchange, both agents' balances are updated simultaneously — a positive change in the cooperator's balance is accompanied by an immediate and opposite change in the counterpart's balance (Figure 1). This reflects the tangible transfer of money that is contingent upon the completion of a cooperative action and binds both agents involved in the exchange.

The second difference is *control*. Reputation is largely beyond personal control, as it is determined by the judgments of others, whereas in a monetary mechanism agents voluntarily agree to an exchange (Frean & Marsland, 2023). This translates into a key difference when agents defect: in the monetary mechanism, agents' balances remain unchanged and do not affect their ability to secure future cooperation from others who value money. In addition, the actions of agents following other strategies do not affect money balances — only agents following the monetary exchange strategy affect their distribution within the population. For example, direct reciprocators make cooperation decisions based on factors independent of their partner's money balances, leaving those balances unaffected. In contrast, reputation is dynamically updated after each action taken by each agent, reflecting a more passive and latent system of judgment over which individuals have little to no control.

Interdependence and control combine to produce the typical *quid pro quo* conditionality observed in everyday monetary transactions (Ostroy, J. M. & Starr, R. M., 1990): tokens change hands only by mutual consent, at the same time as products are given away, and without creating any further obligations between the transactors. These features create a circulation of tokens that binds agents through spot, pairwise, interrelated actions, and helps to promote cooperation. However, our model also shows that the stability of the monetary circuit depends on a delicate balance between the total number of tokens in circulation and the frequency of cooperative actions taken by agents. The main threat to this stability arises when defectors have a significant stock of tokens. Such a stock of tokens allows them to masquerade as a trustworthy party with a history of extensive cooperation and thus continuously benefit from the cooperative actions of agents following the monetary exchange strategy without reciprocating. In other words, an oversupply of money in the system disrupts the information signal that tokens carry about past cooperative behavior. These findings are consistent with empirical observations of money oversupply (M. Friedman, 1994) and recent experimental studies on the relationship between tokens and cooperation (Bigoni et al., 2020). Moreover, they shift the issue of free-riding and defection from interpersonal dynamics to higher-order institutional concerns such as inflation, counterfeiting, and inequality. While the monetary strategy can effectively solve interpersonal cooperation problems by excluding defectors from the circuit, it simultaneously shifts the locus of trust to the institutional framework that must support the proper functioning of the monetary circuit itself (Giddens, 1990), especially in situations where the returns to cooperation are relatively low.

Our findings raise interesting questions for future research. We have suggested that the monetary exchange mechanism allows agents to offload the cognitive effort required to keep track of past interactions through personal memory or in-group reputation onto the tokens themselves, which act as a synthetic indicator of past cooperation. In this sense money, whether in the form of tangible objects or virtual records, could be viewed as a distinctive numerical correlating device (Bowles & Gintis, 2013; Guala, 2020) or social cognitive artifact (Aoki, 2011) that facilitates cooperation through the shared social value attributed to tokens. In other words, money provides a way to 'unburden' (Guala, 2020) individuals of the cognitive effort required to keep track of multiple personal interactions. At the same time, while the monetary strategy is cognitively simple for each individual agent — it only requires checking whether someone has a token — it is *relationally* complex. A successful exchange depends on both parties agreeing on the value of the token, verifying its authenticity, counting it, and ensuring that cooperators receive what they are due through enforceable arrangements.

The Janus-faced nature of money invites reflection on its possible relevance to two famous hypotheses in social evolution, namely the 'social brain hypothesis' (Humphrey, 1976; Dunbar, 1998, 2016) and the 'cultural brain hypothesis' (Muthukrishna et al., 2018). The former emphasizes, in particular, the role of social group size and complex social relationships in creating positive selection pressures for sophisticated social cognition in humans. However, cooperation-signaling artifacts, such as tokens and money, compensate precisely for human cognitive limitations in remembering and tracking past personal relationships, thus solving cooperation dilemmas by socially disembedding cooperation. This is even more compelling when we consider that money is more easily transmitted to offspring than any other benefit, such as reputation and status, thus extending the social disembedding of cooperation benefits even across generations (Ganßmann, 2013).

On the other hand, while money promotes generalized reciprocity across group boundaries, it is primarily a sophisticated cognitive artifact that requires symbolic abstractions that are likely to exist only in a favorable social environment, including a predictable social context of rules and trust. Indeed, the cognitive constraints on social relations that money removes are shifted to high cognitive demands in the form of rules and social learning associated with the cultural environment. This suggests that research on cultural evolution that links the successful trajectory of human societies to the co-evolution of cognition, learning, and complex symbolic artifacts that fostered sophisticated forms of cooperation (Creanza et al., 2017), should also consider the role of money and its institutional regulation (Ingham, 2004; Ganßmann, 2013; Dodd, 2014; Orléan, 2020; Maurer, 2020).

**Acknowledgements:** EF and TA acknowledge financial Support from FCT – Fundação para a Ciência e Tecnologia (Portugal), under Projects UIDB/04521/2020, UI/BD/151563/2021, and UIDB/05069/2020. FR acknowledges support from FAiR, an EU funded project (Grant agreement ID: 101094828). FS acknowledges support from a PRIN-MUR (Progetti di Rilevante Interesse Nazionale – Italian Ministry of University and Research) grant (Grant Number: 202297CKET_002 "Algolit").

**Declarations of Interest**: none

**CRediT author statement:** ECF and FR: conceptualization, methodology, software, validation, formal analysis, writing, visualization. TVA: con ceptualization, methodology. FS: conceptualization, methodology, writing.

**References**

Aoki, M. (2011). Institutions as cognitive media between strategic interactions and individual beliefs. *Journal of Economic Behavior & Organization*, *79*(1), 20–34. https://doi.org/10.1016/j.jebo.2011.01.025
Axelrod, R. (2006). *The Evolution of Cooperation* (Rev. ed). Basic Books.
Axelrod, R., & Hamilton, W. D. (1981). The evolution of cooperation. Science, 211(4489), 1390-1396.
Bigoni, M., Camera, G., & Casari, M. (2019). Partners or Strangers? Cooperation, Monetary Trade, and the Choice of Scale of Interaction. *American Economic Journal: Microeconomics*, *11*(2), 195–227.
Bigoni, M., Camera, G., & Casari, M. (2020). Money is more than memory. *Journal of Monetary Economics*, *110*, 99–115. https://doi.org/10.1016/j.jmoneco.2019.01.002
Bowles, S., & Gintis, H. (2013). *A cooperative species: Human reciprocity and its evolution* (1. paperback print). Princeton Univ. Press.
Camera, G., Casari, M., & Bigoni, M. (2013). Money and trust among strangers. *Proceedings of the National Academy of Sciences*, *110*(37), 14889–14893. https://doi.org/10.1073/pnas.1301888110
Cooper, G. A., & West, S. A. (2018). Division of labour and the evolution of extreme specialization. *Nature Ecology & Evolution*, *2*(7), 1161–1167. https://doi.org/10.1038/s41559-018-0564-9
Creanza, N., Kolodny, O., & Feldman, M. W. (2017). Cultural evolutionary theory: How culture evolves and why it matters. *Proceedings of the National Academy of Sciences*, *114*(30), 7782–7789. https://doi.org/10.1073/pnas.1620732114
Dawkins, R. (1981). *The selfish gene* (Repr. with corr). Oxford Univ. Pr.
Demps, K., & Winterhalder, B. (2019). "Every Tradesman Must Also Be a Merchant": Behavioral Ecology and Household-Level Production for Barter and Trade in Premodern Economies. *Journal of Archaeological Research*, *27*(1), 49–90. https://doi.org/10.1007/s10814-018-9118-6
Dodd, N. (2014). *The social life of money*. Princeton University Press.

Dunbar, R. I. M. (1998). The social brain hypothesis. *Evolutionary Anthropology: Issues, News, and Reviews*, *6*(5), 178–190. https://doi.org/10.1002/(SICI)1520-6505(1998)6:5<178::AID-EVAN5>3.0.CO;2-8

Dunbar, R. I. M. (2016). The Social Brain Hypothesis and Human Evolution. In R. I. M. Dunbar, *Oxford Research Encyclopedia of Psychology*. Oxford University Press. https://doi.org/10.1093/acrefore/9780190236557.013.44

Durkheim, E. (2023). The Division of Labour in Society. In *Social Theory Re-Wired* (3rd ed.). Routledge.

Fauvelle, M. (2025). The Trade Theory of Money: External Exchange and the Origins of Money. *Journal of Archaeological Method and Theory*, *32*(1), 23. https://doi.org/10.1007/s10816-025-09694-9

Frean, M., & Marsland, S. (2023). Score-mediated mutual consent and indirect reciprocity. *Proceedings of the National Academy of Sciences*, *120*(23), e2302107120. https://doi.org/10.1073/pnas.2302107120

Friedman, J. W. (1971). A Non-cooperative Equilibrium for Supergames. *The Review of Economic Studies*, *38*(1), 1. https://doi.org/10.2307/2296617

Friedman, M. (1994). *Money mischief: Episodes in monetary history* (1st Harvest ed). Harcourt Brace & Co.

Gambetta, D. (2011). *Codes of the Underworld: How Criminals Communicate* (1st ed). Princeton University Press.

Ganßmann, H. (2013). *Doing money: Elementary monetary theory from a sociological standpoint* (First issued in paperback). Routledge, Taylor & Francis Group.

Giddens, A. (1990). *The Consequences of modernity*. Polity Press.

Guala, F. (2020). Money as an Institution and Money as an Object. *Journal of Social Ontology*, *6*(2), 265–279. https://doi.org/10.1515/jso-2020-0028

Hamilton, W. D. (1964). The genetical evolution of social behaviour. II. *Journal of Theoretical Biology*, *7*(1), 17–52. https://doi.org/10.1016/0022-5193(64)90039-6

Haour, A., & Moffett, A. (2023). Global Connections and Connected Communities in the African Past: Stories from Cowrie Shells. *African Archaeological Review*, *40*(3), 545–553. https://doi.org/10.1007/s10437-023-09546-5

Hart, K. (2000). *The memory bank: Money in an unequal world* (1. publ). Profile Books.

Harwick, C. (2023). Money's mutation of the modern moral mind: The Simmel hypothesis and the cultural evolution of WEIRDness. *Journal of Evolutionary Economics*, *33*(5), 1571–1592. https://doi.org/10.1007/s00191-023-00844-4

Hauert, C., & Doebeli, M. (2004). Spatial structure often inhibits the evolution of cooperation in the snowdrift game. *Nature*, *428*(6983), 643–646. https://doi.org/10.1038/nature02360

Hayek, F. A. von, & Bartley, W. W. (2000). The collected works of F. A. Hayek. 1: The fatal conceit: the errors of socialism / ed. by W. W. Bartley III (Repr.). Univ. of Chicago Press.

Hilbe, C., Martinez-Vaquero, L. A., Chatterjee, K., & Nowak, M. A. (2017). Memory-n strategies of direct reciprocity. *Proceedings of the National Academy of Sciences*, *114*(18), 4715–4720. https://doi.org/10.1073/pnas.1621239114

Hudson, M. (2020). Origins of Money and Interest: Palatial Credit, Not Barter. In S. Battilossi, Y. Cassis, & K. Yago (Eds.), *Handbook of the History of Money and Currency* (pp. 45–65). Springer. https://doi.org/10.1007/978-981-13-0596-2_1

Humphrey, N. K. (1976). The social function of intellect. In *Growing points in ethology* (pp. 303–317). Cambridge University Press.

Ingham, G. (2004). *The nature of money*. Polity.

Iyer, S., & Killingback, T. (2020). Evolution of Cooperation in Social Dilemmas with Assortative Interactions. *Games*, *11*(4), Article 4. https://doi.org/10.3390/g11040041

Jaeggi, A. V., Hooper, P. L., Beheim, B. A., Kaplan, H., & Gurven, M. (2016). Reciprocal Exchange Patterned by Market Forces Helps Explain Cooperation in a Small-Scale Society. *Current Biology*, *26*(16), 2180–2187. https://doi.org/10.1016/j.cub.2016.06.019

Kocherlakota, N. R. (1998). Money Is Memory. *Journal of Economic Theory*, *81*(2), 232–251. https://doi.org/10.1006/jeth.1997.2357

Kuijpers, M. H. G., & Popa, C. N. (2021). The origins of money: Calculation of similarity indexes demonstrates the earliest development of commodity money in prehistoric Central Europe. *PLOS ONE*, *16*(1), e0240462. https://doi.org/10.1371/journal.pone.0240462

Maurer, B. (2020). Primitive and Nonmetallic Money. In S. Battilossi, Y. Cassis, & K. Yago (Eds.), *Handbook of the History of Money and Currency* (pp. 87–104). Springer. https://doi.org/10.1007/978-981-13-0596-2_2

Muthukrishna, M., Doebeli, M., Chudek, M., & Henrich, J. (2018). The Cultural Brain Hypothesis: How culture drives brain expansion, sociality, and life history. *PLoS Computational Biology*, *14*(11), e1006504. https://doi.org/10.1371/journal.pcbi.1006504

Nax, H. H., & Rigos, A. (2016). Assortativity evolving from social dilemmas. *Journal of Theoretical Biology*, *395*, 194–203. https://doi.org/10.1016/j.jtbi.2016.01.032

Nirjhor, M. S. A., & Nakamaru, M. (2023). The evolution of cooperation in the unidirectional linear division of labour of finite roles. *Royal Society Open Science*, *10*(3), 220856. https://doi.org/10.1098/rsos.220856

Nowak, M. A. (2006). Five Rules for the Evolution of Cooperation. *Science*, *314*(5805), 1560–1563. https://doi.org/10.1126/science.1133755

Nowak, M. A., & Sigmund, K. (1992). Tit for tat in heterogeneous populations. *Nature*, *355*(6357), 250–253. https://doi.org/10.1038/355250a0

Nowak, M. A., & Sigmund, K. (1998). Evolution of indirect reciprocity by image scoring. *Nature*, *393*(6685), 573–577. https://doi.org/10.1038/31225

Nowak, M. A., & Sigmund, K. (2005). Evolution of indirect reciprocity. *Nature*, *437*(7063), Article 7063. https://doi.org/10.1038/nature04131

Okada, I. (2020). A Review of Theoretical Studies on Indirect Reciprocity. *Games*, *11*(3), Article 3. https://doi.org/10.3390/g11030027

Orléan, A. (2020). Money: Instrument of Exchange or Social Institution of Value? In P. Alary, J. Blanc, L. Desmedt, & B. Théret (Eds.), *Institutionalist Theories of Money: An Anthology of the French School* (pp. 239–264). Springer International Publishing. https://doi.org/10.1007/978-3-030-59483-1_8

Ostroy, J. (1973). The Information Efficiency of Monetary Exchange. *American Economic Review*, *63*, 597–610.

Ostroy, J. M., & Starr, R. M. (1990). The transactions role of money. Handbook of monetary economics, 1, 3-62.

Panchanathan, K., & Boyd, R. (2004). Indirect reciprocity can stabilize cooperation without the second-order free rider problem. *Nature*, *432*(7016), 499–502. https://doi.org/10.1038/nature02978

Pisor, A. C., & Gurven, M. (2018). When to diversify, and with whom? Choosing partners among out-group strangers in lowland Bolivia. *Evolution and Human Behavior*, *39*(1), 30–39. https://doi.org/10.1016/j.evolhumbehav.2017.09.003

Powers, S. T., van Schaik, C. P., & Lehmann, L. (2021). Cooperation in large-scale human societies—What, if anything, makes it unique, and how did it evolve? *Evolutionary Anthropology: Issues, News, and Reviews*, *30*(4), 280–293. https://doi.org/10.1002/evan.21909

Redhead, D., Gervais, M., Kajokaite, K., Koster, J., Hurtado Manyoma, A., Hurtado Manyoma, D., McElreath, R., & Ross, C. T. (2024). Evidence of direct and indirect reciprocity in network-structured economic games. *Communications Psychology*, *2*(1), 44. https://doi.org/10.1038/s44271-024-00098-1

Richter, T., Garrard, A. N., Allock, S., & Maher, L. A. (2011). Interaction before Agriculture: Exchanging Material and Sharing Knowledge in the Final Pleistocene Levant. *Cambridge Archaeological Journal*, *21*(1), 95–114. https://doi.org/10.1017/S0959774311000060

Ridout-Sharpe, J. (2015). Changing lifestyles in the northern Levant: Late Epipalaeolithic and early Neolithic shells from Tell Abu Hureyra. *Quaternary International*, *390*, 102–116. https://doi.org/10.1016/j.quaint.2015.11.041

Riolo, R. L., Cohen, M. D., & Axelrod, R. (2001). Evolution of cooperation without reciprocity. *Nature*, *414*(6862), 441–443. https://doi.org/10.1038/35106555

Robinson, E. J. H., & Barker, J. L. (2017). Inter-group cooperation in humans and other animals. *Biology Letters*, *13*(3), 20160793. https://doi.org/10.1098/rsbl.2016.0793

Rossetti, C. S. L., & Hilbe, C. (2024). Direct reciprocity among humans. *Ethology*, *130*(4), e13407. https://doi.org/10.1111/eth.13407

Schmid, L., Chatterjee, K., Hilbe, C., & Nowak, M. A. (2021). A unified framework of direct and indirect reciprocity. *Nature Human Behaviour*, *5*(10), 1292–1302. https://doi.org/10.1038/s41562-021-01114-8

Searle, J. R. (2005). What is an institution? *Journal of Institutional Economics*, *1*(1), 1–22. https://doi.org/10.1017/S1744137405000020

Shin, J., Price, M. H., Wolpert, D. H., Shimao, H., Tracey, B., & Kohler, T. A. (2020). Scale and information-processing thresholds in Holocene social evolution. *Nature Communications*, *11*(1), 2394. https://doi.org/10.1038/s41467-020-16035-9

Simmel, G. (2011). *The philosophy of money*. Routledge.

Singh, M., & Glowacki, L. (2022). Human social organization during the Late Pleistocene: Beyond the nomadic-egalitarian model. *Evolution and Human Behavior*, *43*(5), 418–431. https://doi.org/10.1016/j.evolhumbehav.2022.07.003

Smaldino, P. E., Flamson, T. J., & McElreath, R. (2018). The Evolution of Covert Signaling. *Scientific Reports*, *8*(1), 4905. https://doi.org/10.1038/s41598-018-22926-1

Tkadlec, J., Hilbe, C., & Nowak, M. A. (2023). Mutation enhances cooperation in direct reciprocity. *Proceedings of the National Academy of Sciences*, *120*(20), e2221080120. https://doi.org/10.1073/pnas.2221080120

Wilensky, U. (1999). *NetLogo*. Center for Connected Learning and Computer-Based Modeling, Northwestern University, Evanston, IL. http://ccl.northwestern.edu/netlogo/

Xie, G., Chen, X., Meng, Z., Wu, Y., & Higham, C. (2025). Complex hunter-gatherers and first farmers in southern China. *Archaeological Research in Asia*, *42*, 100615. https://doi.org/10.1016/j.ara.2025.100615